# JOSÉ MONTEIRO DA ROCHA (1734-1819) AND HIS WORK OF 1782 ON THE DETERMINATION OF COMET ORBITS

**AUTHORS:** Fernando B. Figueiredo1 and João M. Fernandes2

1 fernandobfigueiredo@gmail.com or bandeira@mat.uc.pt

2 jmfernan@mat.uc.pt

**AFFILIATIONS:** CITEUC, University of Coimbra http://citeuc.pt/index.php/en/

**ADRESS:** CITEUC - Centro de Investigação da Terra e do Espaço da Universidade de Coimbra, Observatório Geofísico e Astronómico. Santa Clara, 3040 - 004 Coimbra

## ABSTRACT

In 1782 José Monteiro da Rocha, astronomer and professor of the University of Coimbra, presented in a public session of the Royal Academy of Sciences of Lisbon a memoir on the problem of the determination of the comets' orbits. Only in 1799, the '*Determinação das Orbitas dos Cometas*' [Determination of the orbits of comets] would be published in the Academy's memoires. In that work, Monteiro da Rocha presents a method for solving the problem of the determination of the parabolic orbit of a comet from three observations. Monteiro da Rocha's method is essentially the same method proposed by Olbers and published under von Zach's sponsorship two years before, in 1797. To have been written and published in Portuguese was certainly a hindrance for its dissemination among the international astronomical community. In this article, we intend to present Monteiro da Rocha's method and trying to see to what extent Gomes Teixeira's assertion (Teixeira 1934) that Monteiro da Rocha and Olbers must figure together in the history of astronomy, as the first inventors of a practical and easy method for the determination of parabolic orbits of comets, is justified.

**Keywords:** $18^{th}$-century, comets' orbits determination, parabolic orbits, Monteiro da Rocha, Olbers

## 1 INTRODUCTION

Throughout the ages, the comets have always been seen with a mixture of awe and fear. Their rare visits, the variety of directions in which they appeared relative to the Sun and Moon, some in direct motion and others in the retrograde movement. The enormous array of shapes and sizes of their tails, the different glitters with which they appeared, being even visible in broad daylight, reinforced their image of extravagant and mysterious beings. Also, their movements, ranging from slow comets to rapid ones, made them more mysterious and incognito. First looked in the Aristotelian theory as some atmospheric phenomena pass gradually, after the observations of Tycho Brahe (1546-1601) of the comet of 1577, to be regarded as heavenly entity obeying to Newtonian laws. But their physical nature and trajectories remain a mystery for a long time.

The question of the determination of the orbits of comets was one of the most challenging astronomical problems of the 18th-century and had been occupying astronomers and mathematicians for a long time. Attempts of a more mathematical nature can be found from the time Newton (1643-1727) published a rather ingenious method in the *Principia* (1687). That geometrical method was difficult to apply in practice (Kriloff, 1925). Nevertheless, it was using Newton's method, with minor adaptations, that Halley (1656-1742) could determine the orbits of 24 comets, concluding that the 1531, 1607 and 1682 comets displayed identical orbits (Halley, 1705). That led him to suggest that those sightings were in fact to successive and periodical passages of the same comet, which would return in 1758[i]. After Halley's work, virtually nothing of interest was published on the subject. The exception goes to Bradley (1693-1762) who discovered the parabolic orbit of the comet of the year 1723, and Bouguer (1698-1758) who calculated the parabolic orbit of the comet of the year 1729. In the 1740s, due to the significant development of mathematical tools that allowed new analytical approaches, the comets' orbits computing issue took a new breath. At that time, we must highlight the works and publications of Struyck (1686-1769) in 1740, Maraldi (1709-1788) in 1743, Zanotti (1709-1782), Lemonnier (1715-1799), Douwes (1713-1773), and Lacaille (1713-1762) in the year 1746. About Lacaille's method, Delambre (1749-1822) will write: "Lacaille a le premier rendu le problème intelligible. Sa méthode, publiée en 1746, a été longtemps suivie par les astronomes" (Delambre, 1827: xxj). In 1783 Pingré (1711-1796) published a two-volume work, *Cometógraphie au Traité Historique et théorique dès Comètes* (1783-84), where he compared and criticized several methods that had been proposed at different times. He also presents a re-formulation of the Lacaille's method, which thenceforth became known as the Pingré's method and that will be a widely used tool to calculate the comets orbital parameters. Nevertheless, all those methods were not much more than ingenious variations and improvements of Newton's semi-graphic method.

The first analytical method that did not make use of geometrical considerations was presented by Euler (1707-1783) in 1744, in the *Theoria motuum planetarum et cometarum*. The Euler's method makes use of three observations close to each other, considering that the heliocentric distance of the comet at the moment of the second observation is known with a reasonable approximation level. In his book, Euler also presented a set of fundamental theorems for the calculation of the comets orbits. One of them is the famous Euler's equation, which establishes a relationship between the time elapsed between two instants ($t_1$ and $t_2$), the two Sun-comet vector rays ($r_1$ and $r_2$) and the arc path (*s*) described by the comet in that interval of time. Lambert (1728-1777) will provide a practical solution to that equation in his 1761 and 1771 papers, thus becoming known by the Euler-Lambert's equation (Marsden, 1995: 182):

$$6k(t_2 - t_1) = (r_1 + r_2 + s)^{3/2} \pm (r_1 + r_2 - s)^{3/2}, \text{ being k the Gauss constant}[ii]$$

(Equation 1)

In the years that followed, other leading mathematicians and astronomers devoted themselves to the problem. In 1774, Boscovich (1711-1787) developed a method that also made use of three observations, published under the title *De orbitus cometarum determinandis*, in the Mémoires of the Académie Royale des Sciences[iii]. The Boscovich's work was heavily criticised by the Laplace (1749-1827), who found it analytically wrong. The method neglects the 2nd order approximations, both for the curvature as for the changes in comet velocity, but on the other hand makes use of these quantities to calculate the supposed positions of the comet, through the value of the 2nd derivative of the latitude and longitude. Laplace's attack was such that the Paris Academy saw itself obliged to establish a scientific commission to arbitrate the question. The committee would give reason to Laplace (Gillispie 1997: 97). In 1778 Lagrange (1736-1813) presented new contributions towards a better resolution of the problem, and two years later, in 1780, Laplace will give an entirely new approach (Marsden, 1995:183 and Gillispie, 1997:298).

In contrast to others previously methods, requiring three observations relatively close to each other, Laplace's method could make use of observations separated by about 30 to 40 degrees of each other. The precision of the method improves indeed if it uses a more extensive set of observations, and sometimes is genuinely unsatisfactory when only three observations are available. Laplace demonstrated that was possible to calculate the position vector (heliocentric vector) and the velocity vector of the comet and use these results to calculate its orbital elements. Pingré considered the Laplace's method to be the best of all. However, due to the calculation effort required to compute the derivatives, its use has not been widely adopted. Hoeffer (1873: 588) will write:

> "La méthode de Laplace exigeait beaucoup de calculs préparatoires, tellement fastidieux, que les calculateurs, à bout de patiente, abandonnaient quelquefois leur travail, sans avoir obtenu aucun résultat satisfaisant"[iv].

Despite the difficulty of Laplace's method, its fame was such that obscured the contribution of another French mathematician and astronomer Dionis du Séjour (1734-1794), who presented his method to the Paris Academy in 1779[v].

In fact, all these analytical methods were very tedious and complicated. The magnitude of the task was such that in 1772 the Berlin Academy of Sciences proposed a prize to be awarded in 1774 for those who proposed an easy method to calculate a parabolic orbit of a comet upon three observations,

> Il s'agit de perfectionner les méthodes qu'on emploie pour calculer les orbites des Comètes d'après les Observations; de donner surtout les formules générales & rigoureuses qui renferment la solution du Problème où il s'agit de déterminer l'orbite parabolique d'une Comète par le moyen de trois observations, & d'en faire voir l'usage pour résoudre ce probleme de la manière la plus simple & la plus exacte. (Nouveaux Mémoires, 1779: 13).

For lack of competitors, the Berlin Academy extends the deadline of the competition to 1778. The prize was attributed to Condorcet (1743-1794) and Tempelhoff (1738-1808). Johann Friedrich Hennert (1733-1813), who was also a contestant, receives an 'Accessit'. The works of the three men were published in 1780, *Dissertations sur la Théorie des Comètes qui ont concouru au prix proposé par l'Académie Royale des Sciences et Belles-Lettres de Prusse* (Utrecht, 1780).

Nevertheless, it was the German astronomer Olbers who will receive the past credit as the inventor of a simple and easily applicable method with a work published by von Zach (1754-1832), in 1797 – *Abhandlung über die leichteste und bequemste Methode die Bahn eines Cometen aus einigen Beobachtungen zu berechnen*[vi]. In the preface, von Zach wrote:

> I hope no excuse is needed to see this excellent work in print. Instead, the editor wishes to have gained the thanks of all astronomers and lovers of the stellar arts, for having placed such a serious, useful and understandable work on calculating the orbit of comets into their hands. (Olbers, 1797: preface)[vii]

Von Zach's expectations were justified. The method was simple and required a relatively small calculation effort, and so quickly became a tool widely used by astronomers. Olbers's method has received elated compliments from astronomers over time[viii]. As Marsden (1995: 184) points out the method was the pinnacle of orbit-determination during the 18[th]-century, that with a fair amount of justification Olbers gave the title, *"Treatise on the easiest and most convenient method of computing the path of a comet."* A method that was still often used in the early 20[th]-century[ix]. This method will in fact put Olbers' name in history as being the first to discover an easy-to-use computation process of relatively small computational effort for the determination of the parabolic orbits of comets. But maybe the story has something else to say.

Perhaps was not Olbers who was the first to propose such an easy method. In fact, it was the Portuguese mathematician and astronomer José Monteiro da Rocha (1734-1819) who was the first to propose such a method in 1782.

In January 27 of 1782 in a public session of the Royal Academy of Sciences of Lisbon, Monteiro da Rocha presented a work entitled *'Determinação da Orbita dos Cometas'* (Determination of the orbit of comets) where he presents a very similar method to the one that will be proposed circa 16 years later by the German astronomer. Unfortunately, it will be published only in 1799, in the Lisbon Academy Memoires (Rocha, 1799). With that delay, Monteiro da Rocha will have lost to Olbers the priority of being the first to publish a simple and easy method for the determination of the parabolic orbit of a comet through three observations. Monteiro da Rocha also had the intention to publish a second memoire on the problem of the elliptical orbits (Rocha 1799: 405), but as far we know he had never done it[x].

The problem to calculate elliptical orbits was much more difficult. The astronomers used trial and error techniques to find the elliptical orbit of a comet. First, it was estimated a parabola and then tried to accommodate an ellipse that best satisfied the observations. The problem of the elliptical orbits would be solved by Gauss (1777-1855) in 1808 (for Gauss method see Dubois, 1865: 556-580). However, the problem of calculating the orbits remains difficult for many years, in 1828 David Milne wrote:

> And when even astronomers of the present day, with all the advantages of improved science, find so many discrepancies in their calculations, the determination of a comet's orbit may justly be considered one of the most complicated problems in astronomy. (Milne, 1828: 55-56).

## 2 JOSÉ MONTEIRO DA ROCHA (1734-1819): BIOGRAPHICAL NOTES

José Monteiro da Rocha is a crucial figure of the 18$^{th}$-century Portuguese science[xi]. He was born on June 25, 1734, in a small town in the north of Portugal. In 1752 he joined the Society of Jesus and gone to Brazil, where he studied at the famous Jesuit school, Colégio de São Salvador da Bahia. In 1759, following the expulsion of the Jesuits from Portugal by Pombal (1699-1782), the all-powerful minister of King José I (1714-1777), Monteiro da Rocha left the religious order, becoming a grammar teacher in the new state schools that are then created. In Brazil, Monteiro da Rocha observes the passage of comet Halley through the skies of Bahia. He made observations between 13th of March and late April of 1759, without realizing that it was the famous comet. The comet's return would be a crucial test for Newton's theory and Monteiro da Rocha, a staunch supporter of the Newtonian philosophy, took the opportunity to spread the gravitational theory of the English mathematician, writing a text about the nature and orbits of comets (Rocha, 2000).

The text entitled 'Sistema Physico-matemático dos Cometas' (Physical-mathematical system of Comets), dedicated to the governor of the city of S. Salvador da Bahia, is composed of two parts. In the first one, Monteiro da Rocha discusses the judgments of the *"most celebrated philosophers and mathematicians"* about the nature of the comets and shows that they are *"true celestial bodies as old as rest of the heavenly bodies"*. At a time when the Newtonian philosophy about the nature of comets still suffered resistance to be accepted, this part reveals to be is very interesting. Monteiro da Rocha gives us a detail historical perspective of the different theories and evolution of the scientific knowledge about comets since the classical times to his times. The second one, entitled *"Practical astronomy in order to calculate the motions and the ephemerides of the comets"*, presents some basic concepts about spherical astronomy and basic notions on the kinematics of comets and their orbits. As Eulália dos Santos points out, there is no way of denying that the 'Sistema Physico-matemático dos Cometas', reveals itself as part of a less conservative current of the international scientific thought in Portuguese America, in the second half of the 18[th]-century (Santos 2006, 41). But in no way comparable with the memoire on the parabolic orbits published at the Academy of Sciences of Lisbon some years later in 1799. A real scientific work on one of the leading scientific problems that occupied the astronomical community of the time.

In 1765 Monteiro da Rocha returns to Portugal and attends the University of Coimbra, graduating in canon law. In 1771 due to his friendships with the rector of the University he is called by Pombal to participate in the educational reform project of the University of Coimbra. Monteiro da Rocha will be one of the leading designers of the modern curricula for mathematics and astronomy within the framework of the creation of the new faculty of Mathematics and the Astronomical Observatory. In 1772 he was appointed professor of Physics and Applied Mathematics (1772-1783) and in 1783 of Astronomy (1783-1804). Throughout his long life, he will hold high positions in the service of the state and occupy prominent places in several scientific institutions. In 1780 he was elected member of the Royal Academy of Sciences of Lisbon, and in 1795 he was appointed Director of the Royal Astronomical Observatory of Coimbra University. His contribution will become very important in the future of the scientific activity of the observatory. He was not only responsible for its design and construction supervision, but also for the acquisition of its instruments. He was also vice-rector of the University between the years 1786 and 1804. In 1804 Monteiro da Rocha retired from Coimbra University and moved to Lisbon where he will be appointed of the royal prince Pedro (1798-1834), future king Pedro IV of Portugal and first Emperor of Brazil, and his brothers. Monteiro da Rocha died on 11th December 1819.

Monteiro da Rocha's scientific work covered entirely separate mathematical and astronomical domains (Figueiredo 2011, 2014). At a mathematical level are relevant his translations to Portuguese of the Bézout (1730-1783), Marie (1738-1801) and Bossut (1730-1814) textbooks, as well an exciting range of works in integral calculus and numerical analysis. However, it is on the astronomical field that his scientific work stands out. Beyond the comets' orbits memoire, he wrote several scientific articles on a vast range of astronomical subjects, both theoretical and practical (e.g. longitudes, eclipses, astronomical tables). Some of them were translated and published in France by is former student Manuel Pedro de Melo (1765-1833), who worked with Delambre (1749-1822) at the Paris Observatory (Rocha 1808). Monteiro da Rocha was also the scientific mentor behind the applied mathematical and astronomical methods of the utmost scientific production of the observatory, the *'Efemérides Astronómicas'*, published from 1803 onwards.

**2.2 MONTEIRO DA ROCHA'S WORK ON THE DETERMINATION OF THE ORBITS OF COMETS**

Among the works of Monteiro da Rocha, the Portuguese science historiography frequently cites the paper on the determination of the parabolic orbits of comets as being one of his most significant scientific contributions (see Teixeira, 1934; Carvalho, 1985 and Roque, 2003). There are two main reasons for being so. The first is because, in that paper, Monteiro da Rocha published an easy method for calculating the parabolic orbit of a comet from three observations, a method that he had presented to the Academy of

Sciences of Lisbon, back in 1782. The second, and perhaps the most significant is that method is based on the same principles as Olbers', that dates from 1797.

The Academy of Sciences of Lisbon (henceforth ACL) was created on 24 December 1779. One of its primary purposes was to foster the development of science and technology of the country and to make a useful contribution to the economic and social development of Portugal. So, like as their European counterparts, ACL set up a prize system to promote scientific competition in several fields. Another point made in its regulations was the obligation for the presentation and publication of scientific papers by the members.

Shortly after the ACL was created, Monteiro da Rocha was elected, on 16 January 1780. A few months later, in a letter to Luís António Furtado de Mendonça (1754-1830), Secretary of the Academy, dated from 17 July 1780, Monteiro tells of his intention to collaborate with the ACL. He will send a memoir on the determination of the parabolic orbits of comets, that he had been working on for some time. A crucial question to the astronomical science, says Monteiro da Rocha, in such a way that it had been one of the subjects of competition proposed by the Academy of Berlin in the year of 1772. What better scientific contribution could he make to the newly born academy, which had elected him a member of the exact sciences class, he asks. Monteiro da Rocha was completely unaware of the competition outcome. He did not know that the prize had been extended to 1774 and already awarded in 1778 to Condorcet and Tempelhoff. To Monteiro da Rocha his work on the parabolic orbits of the comets was undoubtedly an excellent contribution that could give a unique prestige to Portuguese science and to the ACL. Monteiro da Rocha emphasises that to the Secretary,

> So far, I do not know if the result of that program is known in Portugal; I will do all diligences to finish my work to prevent in case of meeting myself with some of the results which had been crowned in Berlin. Moreover, to attest any priority of mine in the future, and if that would be also of the interest of our Academy, I will tell you that the main basis of my solution consists of the following three equations [and it follows the equations]. In them, there are no more than three unknowns, which are the distances of the comet to Earth projected on the plane of the ecliptic at the time of the three observations.

This letter with the three fundamental equations, as Monteiro da Rocha refers to them, was read three months later by Luis Furtado de Mendonça in the public academic session of 2 October 1780 and kept in the Academy´s safe. It would take sixteenth months for the work be presented in full to the Academy, as it would happen on 27 January 1782. Moreover, it would be only published, due to budgetary constraints, 17 years later in the second volume of the ACL's mathematical and physical proceedings[xii].

Today, in the ACL archives there is a partial text of the memoire (ACL Ms Azul 1462). It is an only six folios manuscript, undated, concerning only the introduction (paragraphs 1 to 7). But, it does not match the published text. It gives in a footnote, that does not appear in the corresponding printed version, some detailed bibliographical references. And there is no mention of to the intention to study the problem of the elliptical orbits. The footnote shows that Monteiro da Rocha had knowledge about the outcome of the Berlin competition and that he has the book with the winning works of Condorcet, Tempelhoff and Hennert (Dissertations 1780),

> Long after my memoir has been delivered, a book, printed in Utrecht in 1780 arrived in Lisbon. That book contains the dissertations attended in Berlin, one by the Marquis de Condorcet, the other by M. Tempelhoff, and two by M. Hennert. (Rocha, ACL Ms Azul 1462)

Moreover, it also refers to a memoire of Du Séjour[xiii], another from Laplace[xiv], the Boskovich's works[xv] and Pingré's Cométographie,

> The memoirs of the Paris Academy of 1779, printed in 1782, contains a dissertation of M. Du Séjour; and those of 1780, published in 1784, one of M. de Laplace. The Boskovich's new works published in 1785 have in the third volume several pamphlets on the same subject. And M. Pingré, in the second volume of his Cométographie, printed in 1784, makes an extensive compilation of all that he found most useful and practicable in the inquiries of the geometers on this question of the orbits. (Rocha, ACL Ms Azul 1462)

We do not know in what year the 'book printed at Utrecht in 1780' reached the hands of Monteiro da Rocha. Neither the memory of Du Séjour, nor the others. It may have been somewhere between 1784, the date of the publication of the 2nd volume of Pingré's Cometography, and the year 1786, as can be inferred by a letter from Monteiro da Rocha addressed to ACL's Secretary, dated from 26 June 1786, informing that all material was delivered to him,

> I received the memoires and the Book which contains those of Berlin. From what I have already seen of them, I confirmed what I had already conjectured, after seeing the 2nd volume of Pingré's Cometographie. This second volume besides presenting the common method, it also presents Euler, Séjour and Laplace's methods, but does not transcribe the memories of Berlin, except for one of Hennert, that only has the accessit. So, since then, I got the idea that the winning works would be very ingenious and sublime, but not very practical, which was the primary end of the prize program; and this is what they really show. (ACL Ms Azul 1944).

Knowing that the ACL was preparing to start the publication of the memoires of the different Classes. A fact that which will only take place in 1789 and just for the Economic memoires. Monteiro da Rocha requests to the Secretary to send him back all his works and in «especially that of the Comets, which I must resume, to make the second part [eliptical orbits]». And it is still justified that about the problem of the parabolic orbits he had found,

> another method, or two different ones, to solve the fundamental equations, without resorting to the differential formulas, which seems me much more expeditious in practice. And so, I want to replace them, and remake for them the examples: in what is necessary some more time. [letter of 26 June 1786] (Rocha, ACL Ms Azul 1944).

Nothing more is known about this, except that the memoire would be published in 1799 in the 2nd volume of the ACL memoires, 'Memórias de Mathematica e Phisica' (Mathematics and Physics Memoires), that had begun to be released two years earlier in 1797.

If we look closely at the dates, we see that in fact, the work of Monteiro da Rocha preceded Olbers by 15-17 years:

- 1780 (17 July) – a letter from Monteiro da Rocha to the Academy Secretary with the three fundamental equations of the method

- 1782 (27 January) – Monteiro da Rocha's work is presented and read in academic session

- 1786-90 – Monteiro da Rocha has the intention to reformulate his work (there are a series of documents, meager in relevant information, which over these years testify to it)

- 1795-96 – Olbers may already be in possession of his method. In the preface, von Zach, the editor of Olbers work, writes that Olbers had mentioned that he had created a method that he had applied "to the comet discovered by him [Olbers] last year". This comet would be the comet that Olbers had discovered on 31 March 1796. The preface is dated from 16 May 1797, so is reasonable to infer that Olbers' method would be from 1795-96.

- 1797 – publication of Olbers method

- 1799 – publication of Monteiro da Rocha method[xvi]

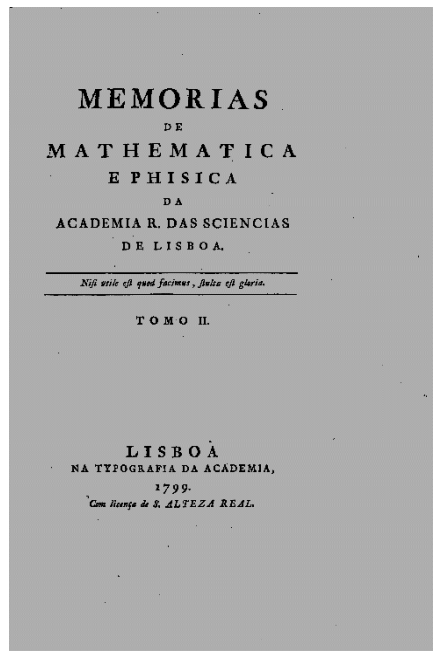

**Figure 1.** The frontispiece of the 2nd volume (1799) of the ACL's memoirs where it was published Monteiro da Rocha work.

It was Duarte Leite (1864-1950), a Portuguese mathematician, professor at University of Porto and diplomat, the first one to carry out a comparative analysis of the equations of the two methods, noticing the similitude between both (Leite 1915),

> La méthode imaginée par l'astronome portugais repose sur les mêmes principes dont s'est servi Olbers, (…) quoique moins commode elle conduit toute fois, surement et sans trop d'effort, au but proposé (…) la formule de l'astronome portugais est, donc, au moins précise que celle d'Olbers (Leite 1915: 66).

In section 4, we will extend the qualitative study of Duarte Leite making a quantitative comparison of the methods, applied to the computation of some comets' parabolic orbits.

### 3 THE ORBIT DETERMINATION PROBLEM

When we talk about the orbital parameters that specify the position and motion in its orbit of a comet or planet, we are referring to 6 parameters: $\Omega$, $\omega$, $I$, $e$, $p$ and $\tau$ (see figure 2). Three of these parameters deal with the orbital orientation about the ecliptic. They are the longitude of the ascending node $\Omega$, which is the angular measurement between the vernal point $\gamma$; the argument of perihelion $\omega$, which is the angular distance between the ascending node and the perihelion point $p$; and the orbital inclination $I$, which is the angle between the orbit plane and the ecliptic plane (if this angle is less than 90º the comet movement is direct, if not it is retrograde). The remaining parameters are connected to the geometrical form of the orbit. The eccentricity $e$ takes the value one if the orbit is parabolic, and $e<1$ if it is elliptic and $e>1$ if it is hyperbolic; the perihelion distance $p$, which is the least distance from the Sun, and $\tau$ is the instant of perihelion passage. If we know these six elements, it is possible to know, without ambiguity, the position of the comet in space. It should be noted that if the orbit is parabolic, the problem reduces to 5 unknown parameters (since the eccentricity is 1).

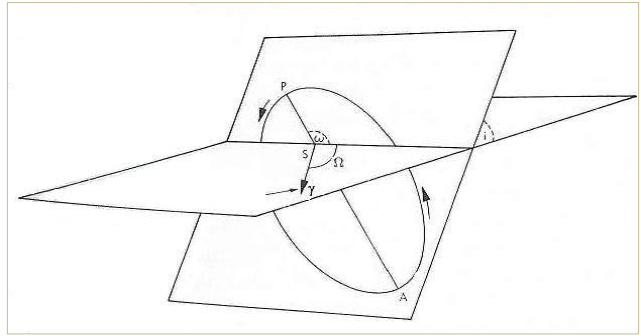

Figure 2: the orbital elements (adapted from Yeomans, p.121)

However, the primary problem is that it is not easy to calculate these comet orbital elements through observations. It is easy to understand that if we know the three distances from the comet to the Sun (or Earth) at the time of the observations, $t_i = (i = 1, 2, 3)$, it would be easy to determine the parabola that fitted to the comet's movement. In fact, the problem would be reduced to a simple geometric question. But we do not know them, observations only give directions (i.e. celestial latitudes and longitudes). How can we get those distances? To do that we need to look more closely at the problem.

Consider the three vectors: $\vec{S_i}$ (Earth-Sun), $\vec{x_i}$ (Earth-comet) and $\vec{r_i}$ (Sun-comet), in such way that: $\vec{x_i} = \vec{S_i} + \vec{r_i}$ (being the subscript i referring to the instant of the ith observation). The $\vec{S_i}$ Earth-Sun vector is completely determined by the astronomical ephemerides, which provides both longitude and latitude of the Sun, as well its distance from the Earth. The $\vec{x_i}$ Earth-comet vector, the geocentric vector, is not completely determined by the observations, they only provide directions, but not magnitudes. It is not possible by direct observation to find the true Earth-comet distance. The 3 geocentric distances of the comet are the 3 unknowns of the problem, that can be formulated in a system of equations that exceeds the number of the unknown by one.

The first equation arises from the condition of the 3 positions (observations) of the comet belonging to a plane that contains the Sun. The second equation follows from the condition of the Sun being the parabola's focus. Comparing the time intervals with the area swept by the vector ray and the related arc chord, we can obtain two more other equations[xvii].

In the 18th-century, due to the state of analytical science at the time, as Olbers (1821: 149-50) writes the most significant difficulty was solving that system of four equations. To overcome such difficulty, astronomers and mathematicians have resorted to simplifying hypotheses and to indirect methods. Two hypotheses were successively considered. One was to assume that the trajectory described by the comet between the time interval $(t_3 - t_1)$ was nearly a straight line covered with constant speed. The other it was to consider that the comet's vector ray at the time of the second observation $(t_2)$ was dividing the chord into two portions that were proportional to the time intervals between the observations. In the late 18th-century, the principal effort of the astronomers was focused on finding another hypothesis that could help to solve the system of equations more easily. Despite many attempts to do so, only Monteiro da Rocha and Olbers were able to. They were the first men who managed to solve in a convenient and easiest way the problem of the determination the orbital parameters of a parabolic comet.

The only essential difference between the method of Monteiro da Rocha and the method of Olbers lies in the approximate relations between the two geocentric distances. Monteiro da Rocha uses an approximate relationship between the middle and the third geocentric distances of the comet, and he uses the Euler-Lambert equation, not in its customary form, but the one obtained by squaring the two members of the theorem. On the other hand, Olbers uses an approximate relation between the two extreme geocentric distances of the comet and a straightforward application of the Euler-Lambert theorem. To Duarte Leite, Monteiro da Rocha's formula is at least as accurate as Olbers's (Duarte Leite 1915: 66, 70).

## 4. THE 'DETERMINAÇÃO DA ORBITA DOS COMETAS' MEMOIRE

Monteiro da Rocha's memoire is organized into 12 sections, in a total of 79 pages. And can be roughly divided into five different parts.

Part 1 is the introduction, where Monteiro da Rocha gives a quick history of the orbit problem and its importance to astronomy. It is very superficial, and it cannot be compared at all with the introduction made by the German. Olbers provides an excellent and extensive history of the problem, describing in detail and contextualizing all the previous contributions, stating that his work is no more than a consequence of a whole body of knowledge: «(…) Euler and Lambert have supposed the same for the orbit of the comet; I have only extended the supposition to that of the earth» (Olbers, 1821-22: 141). Monteiro da Rocha introduction is very intriguing. Indeed, the references and quotes made are scanty. Monteiro da Rocha refers Newton's pioneering work, Euler's work of 1744 and briefly mentions the work of Halley, Lacaille, Pingre and Lalande. About other contributions, he is crushing: «Not taking into account the solutions of Bouguer, Fontaine or others, which never took the necessary step from speculation to execution.» (Rocha, 1799: 404). In fact, whoever reads Monteiro da Rocha's memoire one thing is clear, the previous works from other astronomers have contributed only a little to his own work. An illustrative is the relation between the two vectors, the arc between them and the time that the comet spent to travel it, was of his formulation. Monteiro da Rocha does not refer the Euler-Lambert's theorem.

The part 2, deals with the mathematical and physical analysis of the problem. On one clear way, and step by step, Monteiro da Rocha formulates and establishes the several different equations involved in the problem. It is here that he presents the three fundamental equations of the problem that previously had communicated to Luis Furtado de Mendonça, in the letter of 1780. These three equations establish the relationship between two vector rays $(r_i, r_j \therefore i \neq j = 1, 2, 3)$, the chord of the arc $(k_i)$ described by the comet, and the time $(t_i)$ spent by the comet in that path:

$$R_i(R_i^2 + 3K_i^2) - \sqrt{(R_i^2 - K_i^2)^3} - \Phi t_i^2 = 0, \text{ where } \Phi \text{ is a constant.} \qquad \text{(Equation 2)}$$

This equation is nothing, but the same relation established by Euler-Lambert's theorem (not in its usual form, but as said before in that obtained by squaring the two parts of the theorem). It is interesting to note that Monteiro makes no mention of Euler-Lambert's theorem. Monteiro da Rocha by solving these equations gets the three geocentric distances, which are in fact the real unknowns of the problem ($R$ and $K$ are functions of $x$ and $x''$. Part 2 ends with some considerations on possible methods to solve them. Monteiro da Rocha quickly concludes that it is impossible to solve them directly (due to the high level of the resulting equation), concluding that it is necessary to use indirect methods, namely the method of false position.

In part 3, Monteiro da Rocha establishes the relationship between the three unknown variables (the comet's geocentric distances) at the three instants of observation. That is made doing some physical and geometrical considerations between the arc described by the comet, during the time interval of the 1$^{st}$ and 3$^{rd}$ observations, and the movement of the Earth during that same time interval. A relationship similar to the one that Olbers also formulates in his work,

> We may approach much nearer to the truth by adhering to the supposition, that the chord of comet's orbit is divided by the middle radius in the proportion of the times; and if we assume at the same time that the chord of the Earth's orbit is divided in the same proportion, we shall obtain an approximate solution, which is indirect, but more easy and convenient than could well have been imagined, considering the intricate nature of the problem. (Olbers, 1821-22: 425).

The Monteiro da Rocha equations are:

$$x_1 = h_1 + m_1(x_2 - \chi_1) + \frac{\Theta m_1(q_1 - x_2)}{r_2^3} \qquad \text{(equation 3)}$$

and

$$x_3 = h_2 + m_2(x_2 - \chi_1) + \frac{\Theta m_2(q_2 - x_2)}{r_2^3} \qquad \text{(equation 4)}$$

These both equations combined with one of the fundamental equations (equation 2) gives three different ways of solving the problem approximately. That is, assuming an initial value for $x_2$ we obtain $x_1$ from equation 3. Now making use of equation 2 (to $i = 2$), we have two possible outcomes: or .(Equation is false ($\neq 0$) or is true ($= 0$). If it is false, we make a new assumption for $x_2$ and the process starts again. If .(Equation is true, then we can find $x_3$ from the (equation 4) 4). Through this process of the false position, we can find the three geocentric distances of the comet ($x_1$, $x_2$ and $x_3$) and therefore the orbital elements of the comet (Monteiro da Rocha provides the necessary equations for this purpose).

In part 4 Monteiro da Rocha applies the method to three examples: to a hypothetical comet with a rigorous parabolic orbit; to the comet of the year 1680, and to Halley's Comet of 1759. And iIn the last part of the work, Monteiro da Rocha deals with the inverse problem, the determination in any instant of the position of a comet if its orbital elements are known.

Let see some results and comparisons of Monteiro da Rocha's method.

### 4.1 QUANTITATIVE RESULTS OF MONTEIRO DA ROCHA'S METHOD

In quantitative terms, the results indicate that the solution proposed by Monteiro da Rocha is numerically comparable with that of Olbers. In fact, the results obtained are consistent and with a high degree of precision.

As we have said, one of the examples given by Monteiro da Rocha in his memoire was the determination of the orbit of a hypothetical parabolic comet. Fixing first the orbital parameters of a real parabolic comet ($e = 1$), he calculates 3 hypothetical observations. To that observations, he applies his method to calculate the orbit of the comet, that he compares with what he had previously fixed.

**Table 1:** orbital elements for the hypothetical parabolic comet (this is the 1st example given by Monteiro da Rocha in his paper).

| Orbital elements | hypothetical comet | Monteiro da Rocha's method |
|---|---|---|
| Perihelion distance | 1 | 0,999987 |
| Instant of perihelion passage | 20 March, 12:00:00h | 20 March, 11:57:19h |
| Longitude of the ascending node | 140º | 139º 59' 59'' |
| Longitude of perihelion | 100º | 99º 59' 50'' |
| Inclination of the orbit | 30º | 30º 0' 2'' |

In his second example, Monteiro da Rocha calculates the orbital elements of the Comet 1680.

**Table 2:** orbital elements of the comet of the year 1680. On the left, the results of applying Monteiro da Rocha's method; on the right side, the Encke's calculations (1818) (Watson, 1964: 639).

| Orbital elements | Monteiro da Rocha's method | Encke |
|---|---|---|
| Perihelion distance | 0,006265 | 0,006222 |
| Instant of perihelion passage | 17 Dec. 1680, 23:55:36h | 17 Dec. 1680, 23:46:09h |
| Longitude of the ascending node | 272º 05' 43'' | 272º 09' 29'' |
| Longitude of perihelion | 262º 48' 51'' | 262º 49' 05'' |
| Inclination of the orbit | 60º 45' 41'' | 60º 40' 16'' |

As we can see, the results are practically the same. For example, the difference in the perihelion distance is less than 0.07 %, and for the other parameters, the differences are less than 0.01%. That is even more remarkable, taking into account that the Encke's results are getting from the use of a model for an elliptical orbit.

Monteiro da Rocha also applied his method to Halley's Comet (1759). Monteiro da Rocha uses the observational data from Lalande (1765). The next table presents the results of Monteiro da Rocha, comparing them with the results of Maraldi (1759) and Lalande (1765).

**Table 3:** the orbital elements of Halley's Comet.

| Orbital elements | Monteiro da Rocha | Lalande | Maraldi |
|---|---|---|---|
| Perihelion distance | 0,598770 | 0,584903 | 0,583600 |
| Instant of perihelion passage | 13 March 1759, 11:01:12h | 12 March1759, 13:59:24h | 12 March1759, 12:57:36h |
| Longitude of the ascending node | 54º 9' 43'' | 53º 45' 35'' | 53º 49' 21'' |
| Longitude of perihelion | 299º 38' 21'' | 303º 8' 10'' | 303º 16' 20'' |

| Inclination of the orbit | 17º 27' 55'' | 17º 40' 14'' | 17º 35' 20'' |

As we see, the relative differences are less than 6% (for the orbit inclination) and 1% (for the longitude of perihelion). Moreover, if we compare Monteiro da Rocha's results with some more recent results (using elliptical determination methods), we obtain very good results: 5% for the orbit inclination, and 1% for the longitude of perihelion (see Watson, 1964:640 and Kronk, 1999: 429).

The next table (table 4) lists the results of applying Monteiro da Rocha's method to the first example given by Olbers, the comet of 1769, "partly because the elements of this orbit are so established, and partly because it has been the most frequently employed for an example of other methods." (Olbers, 1821: 141).

**Table 4:** Determination of the orbital elements of the comet of the year 1769 (first example given by Olbers of the application of his own method).

| Orbital elements | Olbers | Monteiro da Rocha |
|---|---|---|
| Perihelion distance | 0,1438005 | 0,108730 |
| Instant of perihelion passage | 7 October 1769, 10:23h | 7 October 1769, 00:23h |
| Longitude of the ascending node | 175º 18' 13'' | 175º 49' 18'' |
| Longitude of perihelion | 149º 59' 31'' | 136º 54' 33'' |
| Inclination of the orbit | 41º 26' 36'' | 43º 16' 31'' |

Comparing these results with Bessel's calculations (1810) using an elliptical model ($e = 0,999249$, so its orbit can be taken as being a parabola): perihelion distance 0,122755; instant of perihelion passage, 8 October 1769 at 02:53h; longitude of the ascending node, 178º 17' 35''; longitude of the perihelion, 149º 7' 21'' and the inclination of the orbit, 40º 44' 02'' (Kronk 1999: 447); the relative differences to Bessel's results of Monteiro da Rocha's and Olbers' methods are identical.

Applying the method of Monteiro da Rocha to the determination of the orbit of the fifth comet seen in 1863, the Comet 1863 (V), and comparing it with the orbital elements computed by James C. Watson (1964: 199-212, 645), we reach practically to the same results.

**Table 5:** The orbital elements of the Comet 1863 (V), using Monteiro da Rocha's method.

| Orbital elements | Watson's calculations | By Monteiro da Rocha's method |
|---|---|---|
| Perihelion distance | 0,771574 | 0,777129 |
| Instant of perihelion passage | 27 Dec. 1863, 13:33:11h | 28 de Dec. 1863, 8:13:14h |
| Longitude of the ascending node | 304º43' 11,5'' | 305º 0' 16,2'' |
| Longitude of perihelion | 60º 23' 17,8'' | 60º 29' 23'' |
| Inclination of the orbit | 64º 31' 27,7'' | 65º 16' 28,5'' |

## 5 FINAL REMARKS

The publication in 1797 of 'The easiest and most convenient method of calculating the orbit of a comet from observations', would put Wilhelm Olbers as the first astronomer who had supplied an easy-to-use method to solve such complex and difficult astronomical problem. All the previous methods required lengthy computations. Olbers's method, with small improvements, was in very general use in the early of the $20^{th}$-century and as Moulton (1970: 259) points out was given in nearly every treatise on the theory of determining orbits. In 1799, two years after Olbers's famous work, the Portuguese astronomer Monteiro da Rocha published a memoir on the same subject. In his academic memoire, Monteiro da Rocha presents an identical analytical method to Olbers'. Curiously, both methods are based on the same principles. The differences between the methods lie in the fact that Monteiro da Rocha makes use of an approximate relation between the geocentric distances of the middle position and the terminal position of the comet, and of the Euler-Lambert equation, not in its habitual form, but one obtained by squaring the two members of the theorem. On the other side, Olbers's method is characterised by the employment of an approximate relation between the geocentric distances of the terminal positions of the comet and the straight application of the theorem of Euler-Lambert. The method of Monteiro da Rocha is also simple and easy as like Olbers' to compute the parabolic orbit of a comet. The Olbers's method which has become classical does not differ substantially from that which is given by Monteiro da Rocha. If we compare both methods, the method of the Portuguese is more exact and accurate than that of the German.

Although the publication of Olbers's work had preceded Monteiro da Rocha's by two years, the timing of the invention of their methods was the reverse. In fact, the work of Monteiro da Rocha Monteiro da was presented and read in the Academy of Sciences of Lisbon in January 17, of 1782. The fact that it was written and published in Portuguese was certainly a factor so that it had not been known beyond national frontiers, thus leading the precedence of the 'invention of a simple and easy method' to have been lost to Olbers. But one thing is sure, the names of Monteiro da Rocha and Olbers must, therefore, appear together in the history of astronomy as the first inventors of a practical and easy method for the determination of the parabolic orbits of comets.

## 6 ACKNOWLEDGES

The authors thank all the support for the conditions provided by CITEUC for the writing of this article. J. Fernandes acknowledges funding from POCH and Portuguese FCT grant SFRH/BSAB/143060/2018. The authors thank Vitor Monteiro for his help in reviewing English.

CITEUC is funded by National Funds through FCT - Foundation for Science and Technology (Projects UID/00611/2020 and UIDP/00611/2020).

## 7 NOTES ON CONTRIBUTORS

Fernando B. Figueiredo is the currently Coordinator of CITEUC. His research interests are the history of mathematics and astronomy in the Enlightenment. More recently his interests have turned to the 19th-century observatory and physical-mathematical sciences (astronomy, geodesy, meteorology, geomagnetism and seismology).

João Fernandes is astronomer and assistant professor in the Mathematics Department at the University of Coimbra. He received a Ph.D. in Astrophysics from the University of Paris in 1996 and the habilitation in Physics/Astrophysics from the University of

Coimbra, in 2014. His research is in solar physics and stellar evolution with interest in history of astronomy, mainly in Portugal.

**8 REFERENCES**

(1779) Nouveaux Mémoires de l'Académie Royale des Sciences et Belles-Lettres. Année MDCCLXXVI, avec l'Histoire pour la méme année,1779. Berlin, George Jacques Decker

(1780) *Dissertations sur la théorie des cométes qui ont concouru au prix proposé para l'Académie Royale des Sciences et Belles Letres de Prusse, pour l'année 1777, & adjugé en 1778.* Publiées avec permission de l'Académie. Utrecht : Barthelemy Wild

(1782) Academia Real das Sciencias. *Gazeta de Lisboa*, Supplemento V, 2 de Fevereiro.

(1801) Monatliche Correspondenz zur Beförderung der Erd- und Himmels-kunde, October, 1801. Gotha

Boscovich RJ (1774) De orbitus cometarum determinandis ope trium observationem parum a se invicem remotarum. Paris

Boscovich RJ (1785) Nouveaux ouvrages de M. l'abbé Boscovich appartenants principalement à l'optique et à l'astronomie en cinq volumes. Venise: Remondini

Broughton P (1985) The first predicted return of comet Halley. *Journal for the History of Astronomy* 16: 123-133

Carvalho R (1985). *A Astronomia em Portugal no século XVIII*. Lisboa: Instituto de Cultura e Língua Portuguesa

Delambre J-B (1827). *Histoire de l'astronomie au dix-huitième siècle.* Paris: Bachelier

Du Séjour APD (1782) Détermination des Orbites des Comètes. HARS, Mémoires de Mathématique & de Physique, année 1779, pp.51-168

Du Séjour, APD (1782) Détermination des Orbites des Comètes. HARS, Mémoires de Mathématique & de Physique, année 1779: 51-168

Dubois E (1865). *Cours d'Astronomie (2$^{nd}$ ed).* Paris: Arthus Bertrand

Fabritius W (1883) Du Séjour und Olbers: ein Beitrag zur Geschichte des Cometenproblems. *Astronomische Nachrichten* 106: 87-94

Figueiredo FB (2005) *A Contribuição de José Monteiro da Rocha para o cálculo das orbitas dos cometas.* M.Phil Thesis, University Nova de Lisboa, PT

Figueiredo FB (2011) *José Monteiro da Rocha e a actividade científica da 'Faculdade de Mathematica' e do 'Real Observatório da Universidade de Coimbra': 1772-1820.* PhD Thesis, University of Coimbra, PT

Figueiredo FB (2014) José Monteiro da Rocha. In Hockey T (Ed), The Biographical Encyclopedia of Astronomers (2nd ed). New York: Springer, pp.513-515

Figueiredo FB (2015) From Paper to Erected Walls: The Astronomical Observatory of Coimbra: 1772–1799. In Raffaele Pisano (ED) *A Bridge between Conceptual Frameworks: Sciences, Society and Technology Studies*. Dordrecht: Springer Netherlands, pp. 155-178

Figueiredo FB and Fernandes JM (2005) *Comparison between Monteiro da Rocha and Wilhelm Olbers' Methods for the determination of the orbits of comets. In:* Afonso J, Santos N, Moitinho A and Agostinho R (eds) *2005: Past Meets Present in Astronomy*

*and Astrophysics. Proceedings of the 15th Portuguese National Meeting.* Singapore: World Scientific Publishing, pp. 85-88

Gillispie CC (1997) *Pierre-Simon Laplace, 1749-1827. A life in Exact Science.* Princeton: Priceton University Press

Halley, E (1705) *Synopsis of the Astronomy of Comets.* London: John Senex

Hoefer F (1873) *Histoire de l'Astronomie.* Paris: Hachette

Kriloff AN (1925) On Sir Isaac Newton's method of determining the parabolic orbit of a comet. *Monthly Notices of the Royal Astronomical Society*, 85(7): 640-656

Kronk GW (1999) *Cometography, a catalog of Comets, volume 1: Ancient-1799.* Cambridge: Cambridge University Press

Lalande JJ (1765) Sur le retour de la comète de 1682, observe en 1759, avec les élémens de son orbit pour cette dernière apparition. Histoire de l'Académie royale des sciences... avec les mémoires de mathématique & de physique... tirez des registres de cette Académie, 1-40

Lalande JJ (1803) *Bibliographie Astronomique, avec l'Histoire de l'Astronomie depuis 1781 jusqu'à 1802.* Paris: l'Imprimerie de la Republique

Laplace PS (1784) Mémoire sur la Détermination des Orbites des Comètes. Mémoires de l'Académie Royale des Sciences de Paris, année 1780, pp.13-72

Leite D (1915) Pour l'Histoire de la Détermination des Orbites Cométaires. *Anais da Academia Politécnica do Porto* X(2): 65-73

Leuschner AO (1913) A short method of determining orbits from three observations, part 1. Publication of the Lick Observatory 7: 3-20

Leuschner AO (1913) Short methods of determining orbits. Second paper, Publication of the Lick Observatory 7: 217-376

Marsden BG (1995) Eighteenth- and nineteenth-century developments in the theory and practice of orbit determination. In: Taton R and Wilson C (Eds) *Planetary astronomy from the Renaissance to the rise of astrophysics. Part 2B: The eighteenth and nineteenth centuries.* Cambridge: Cambridge University Press, pp.181-190

Milne D (1828) *Essay on Comets, which gained the first of Dr. Fellowes's Prizes, proposed to those who had attended The University of Edinburg*. Edinburgh: Adam Black

Moulton FR (1914) *An Introduction to Celestial Mechanics*. New York: MacMillan Company

Olbers W (1797) *Abhandlung über die leichteste und bequemste Methode die Bahn eines Cometen aus einigen Beobachtungen zu berechnen von Wilhelm Olbers.* Weimar: Verlage des industrie-Comptoirs

Olbers W (1820-22) An essay on the easiest and most convenient method of calculating the orbit of a comet from observations, by William Olbers. Weimar, 1797. Translated from German, with notes. *The Quarterly Journal of Science, Literature, and the Arts*, 9(1820): 149-162, 10(1821): 416-426, 11(1821): 177-182, 12(1822): 137-151, 13(1822): 366-385

Rocha JM (1799) Determinação das Orbitas dos Cometas. *Memórias de Mathematica e Phisica da Academia Real das Sciencias de Lisboa* 2:402-479

Rocha JM (1808) *Mémoires sur l'Astronomie Pratique*. Paris: Courcier

Rocha JM (2000). Sistema Físico-Matemático dos Cometas [1760]. In: Camenietzki CZ and Pedrosa FM (Eds). Rio de Janeiro: MAST

Rocha JM [n. d.] Determinação das Orbitas dos Cometas [manuscript]. ACL's Archives, Ms. Azul n.1462.

Santos E (2006) *Dos cometas do nordeste aos thesouros da Amazônia: os jesuítas João Daniel e José Monteiro da Rocha no contexto das Ciências naturais do século XVIII*. PhD Thesis, University of Federal do Paraná, BR

Teixeira FG (1934) *História das Matemáticas em Portugal.* Lisboa: Academia das Ciências de Lisboa

Waff CB (1995) Predicting the mid-eighteenth-century return of Halley's Comet. In: Taton R and Wilson C (Eds) Planetary astronomy from the Renaissance to the rise of astrophysics. Part 2B: The eighteenth and nineteenth centuries. Cambridge: Cambridge University Press, pp. 69-87

Watson JC (1964) *Theoretical astronomy, relating to the motions of the heavenly bodies.* New York: Dover

Yeomans DK (1991) *Comets: a chronological history of observation, science, myth, and folklore.* New York: John Wiley & Sons.

Yvon Villarceau AJ (1857) Détermination des orbites des Planètes et des Comètes. Annales de l'Observatoire Imperial de Paris 3: 1-197

---

[i] Halley had suggested that in order to make a rigorous calculation of the return of the comet it would be necessary to study the gravitational influence of the planet Jupiter but, by realizing the complexity of the necessary calculations, it would remain by mere suggestion. It will be Clairaut (1713-1765), with the help of Lalande (1732-1807) and Nicole-Reine Lepaute (1723-1788), who recovering Halley's idea throws themselves into the calculation of the comet's return. The presentation of the results of that titanic effort was made to the Académie Royal des Sciences on November 14, 1758. They fail for 33 days. The comet reached its perihelion on March 13, 1759, and Clairaut's prediction advanced with the date of 15 April 1759 (Broughton 1985 and Waff 1995).

[ii] The Gauss gravitational constant (symbol k) is a constant value for the solar system, equal for all planets and comets: k = √ (GMs) ≈0.0172020), where G is the gravitational constant G = 6.67 × 10$^{-11}$N .m²kg$^{−2}$). Their value is also given by the formula: $k = \frac{2\pi}{T}\sqrt{\frac{a^3}{M_s m}}$, so it is possible to determine the period of revolution (T) of a planet of mass (m) given the half-largest axis (a) of its orbit, or vice versa.

[iii] *De orbitus cometarum determinandis [...]*, published in Mémoires de Mathématique et de Physique, présentes à Académie Royale des Sciences par divers Savants (v.6, pp.198-215) – this work is an improvement on his previous works of 1746 and 1749.

[iv] The Laplace method would later undergo major improvements; see Villarceau 1857 and Leuschner 1913.

[v] In 1883 the German astronomer W. Fabritius wrote a paper about the supposed Du Sejour's influence on Olbers work (Fabritius 1883).

[vi] Some years later was translated into English and published by the Royal Institution of London, under Olbers own supervision, 'An essay on the easiest and most convenient method of calculating the orbit of a comet from observations' (Olbers, 1821).

[vii] Free translation of von Zach's preface (Olbers 1797).

[viii] In history there are several astronomers' referring Olbers's work in the most laudatory terms. Laplace wrote the following concerning Olbers' method: *"Ce traité dés comètes est un dés meilleurs qu'un ait faits"* [Lalande 1803:638]; Von Zach, who was to be the publisher of Olbers's work, wrote: *"I hope no excuse is needed to see this excellent work in print. Rather, the editor hopes to have gained the thanks of all astronomers and lovers of the stellar arts, for having placed such a serious, useful and understandable work on calculating the orbit of comets into their hands."* [von Zach's preface in Olbers 1797]. David Milne, a young British astronomer who won Edinburgh University's astronomy-prize with an essay on comets, wrote: «*the method which we have now demonstrated, of discovering what the elements of its orbit are from three observations of the comet, is considered to be one of the simplest and most convenient known to astronomers*» [Milne 1828:68]; in the end of 19$^{th}$-century J. Craig Watson wrote: «*He* [Olbers's method] *has managed to achieve a solution which could be performed with remarkable ease (...). The accuracy of the results obtained by Olbers' method and the facility of its application, caught the attention of Legendre, Ivory, Gauss, and Encke to this subject, and through them the method was extended and generalized...*» [Watson, 1964:4]; and more recently the 20$^{th}$-century astronomer, historian, and specialist in comets, Brian G. Marsden, reinforced this: "*The pinnacle of orbit-determination achievement during the eighteenth century was therefore the publication by the Bremen physician H. W. M. Olbers (1758-1840) in 1797 of an essay to which, with a fair amount of justification, he gave the title Treatise on the easiest and most convenient method of computing the path of a comet."* [Marsden 1995:184]; as Yeomans as well: "*Several complete methods were available in the second half of the eighteenth century, but they all required lengthy computations. What was needed was an easy-to-use method, and just such a technique was finally supplied by Heinrich Wilhelm Matthias Olbers (...) It is a credit to his 1797 work that this technique, modified only slightly, is still in use today."* [Yeomans 1991: 144-145].

[ix] "*This method has not been surpassed for computing parabolic orbits and is in very general use even at the present time. It is given in nearly every treatise on the theory of determining orbits*" (Moulton 1914: 259)

[x] In fact, the work *Determinação das Orbitas dos Cometas*, was labelled as 'first part' and addresses only the problem of the parabolic orbits. The 'second part', wich in the intruduction Monteiro da Rocha informs that he will do, does not appear to have been made, any work of Monteiro da Rocha on elliptical orbits is unknown.

[xi] About Monteiro da Rocha's life, scientific and academic work see Figueiredo 2011 and 2014.

[xii] In the 1780's the ACL starts the publication of its Economic memoires, leaving the publication of the mathematical and physical memoires to the end of 1790's (the first volume of this class comes out in 1797 and second in 1799).

[xiii] Du Séjour 1782.

[xiv] Laplace 1784.

[xv] Boscovich 1774 and 1785.

[xvi] Von Zach refers to Monteiro da Rocha work in the Monatliche Correspondenz (October, 1801), but we don't know if he truly studied it. "Bestimm. der Comtenbahnen von Monteiro da Rocha. Diese Abhandlung ward im Jahr 1782 der Academie vorgelegt. Er hat seine Methode mit Erfolg auf die Cometen von 1759 und 1780 angewandt; allein man hat jetzt kürzere Methoden. Der Verfasser handelt am Ende von der Bestimmung der elliptischen Bahnen." (p.355); as we already notice, and to contrary to what Von Zach says, Monteiro da Rocha don't study the elliptical orbits.

[xvii] If we generalize to $n$ observations, we have $(3n-5)$ equations. $(n-2)$ of these equations arises from the conditions of the Sun being in the plane of the comet's orbit; another set of $(n-2)$ equations are consequence of the parabola's properties. The $(n-1)$ remaining equations result from the relation between times, distances and the related arc chords.